\documentclass[aps,amsmath,superscriptaddress,showpacs,prl,twocolumn,longbibliography]{revtex4-2}
\usepackage{bbm}
\usepackage{bm}
\usepackage{enumerate}
\usepackage{graphicx}
\usepackage[dvips]{epsfig}
\usepackage{epsf}
\usepackage[latin1]{inputenc}
\usepackage{mathtools}
\usepackage{amsmath}
\usepackage{amsfonts}
\usepackage{amssymb}
\usepackage{xcolor}
\usepackage{esint}

\begin{document}
	\title{Spin Hall effect of vorticity}
\author{Edward Schwartz}
\affiliation{Department of Physics and Astronomy and Nebraska Center for Materials and Nanoscience, University of Nebraska, Lincoln, Nebraska 68588, USA}

\author{Hamed Vakili}
\affiliation{Department of Physics and Astronomy and Nebraska Center for Materials and Nanoscience, University of Nebraska, Lincoln, Nebraska 68588, USA}

\author{Moaz Ali}
\affiliation{Department of Physics and Astronomy and Nebraska Center for Materials and Nanoscience, University of Nebraska, Lincoln, Nebraska 68588, USA}

\author{Alexey A. Kovalev}
\affiliation{Department of Physics and Astronomy and Nebraska Center for Materials and Nanoscience, University of Nebraska, Lincoln, Nebraska 68588, USA}
	
\begin{abstract}
Using mapping between topological defects in  an easy-plane magnet and electrical charges, we study interplay between vorticity and spin currents. We demonstrate that the flow of vorticity is accompanied by the transverse spin current generation -- an effect which can be termed as the spin Hall effect of vorticity. We study this effect across the BKT transition and establish the role of dissipation and spin non-conservation in the crossover from spin superfluidity to diffusive spin transport. Our results pave the way for low power computing devices relying on vorticity and spin flows that can propagate over long distances.

\end{abstract}

\maketitle

Experimental realizations of van der Waals (vdW) magnetic materials such as NiPS$_3$, CrCl$_3$, and Fe$_3$GeTe$_2$ show unconventional magnetic behavior~\cite{Jiang2021} which can be of interest for the field of spintronics~\cite{Sierra2021}. As vdW systems can be realized in a two-dimensional form, they possess unique properties and result in unusual physics~\cite{Kim2019,Kim2019-1,BedoyaPinto2021}. Possible applications in atomically thin memory and computing devices~\cite{Ryu2020} further motivate research of spin-orbit torques~\cite{Alghamdi2019,Shin2022}, realizations of skyrmions~\cite{Tong2018}, magnetization switching~\cite{Kao2022}, etc. Achieving low dissipation in devices can often become possible by employing concepts of topology and topological protection~\cite{Gilbert2021}.  
Skyrmions, characterized by a topological charge and proposed as information carriers, have been observed in vdW magnetic materials~\cite{Wu2020,Chakraborty2022}. Two-dimensional magnetic systems with easy-plane anisotropy can host magnetic merons -- magnetic defects characterized by a topological vorticity number~\cite{Augustin2021}.  

Two-dimensional vdW magnets can also be used to explore fundamental questions of magnetism related to magnetic phase transitions~\cite{PhysRevB.99.180502,PhysRevLett.127.037204}. The physics associated with the topological defects can lead to the magnetic Berezinskii-Kosterlitz-Thouless (BKT) transition, which is a topological phase transition~\cite{berezinskii1971,Kosterlitz1973}.
Bound topological defects appear below the BKT transition, and they unbind above the BKT transition, thus, leading to various transport signatures~\cite{PhysRevLett.125.237204,PhysRevB.104.L020408,SciPostPhys,PhysRevResearch.4.023236}. The behavior of topological defects can be further mapped to electrodynamics in two dimensions where the vorticity number plays the role of charge while the spin density and current play the role of magnetic and electric fields~\cite{Kosterlitz_1974,RevModPhys.59.1001,PhysRevLett.124.157203,PhysRevB.58.8464}.
The interplay of spin and vorticity currents then results in the crossover~\cite{SciPostPhys,PhysRevResearch.4.023236} from the spin superfluid transport~\cite{PhysRevLett.112.227201,PhysRevB.90.094408,PhysRevLett.115.237201,PhysRevB.95.144432,PhysRevLett.116.117201,PhysRevB.96.134434,PhysRevB.103.144412,PhysRevB.103.L060406,PhysRevB.103.104425,Stepanov2018,Yuan2018} below the BKT transition to the diffusive spin transport above the BKT transition.  

The discovery of the spin Hall effect played an important role in development of spintronics~\cite{RevModPhys.87.1213}. Spin Hall effect and its analogues can be induced by flows of electrons~\cite{Dyakonov1971}, magnons~\cite{PhysRevLett.117.217202,PhysRevLett.117.217203}, phonons~\cite{Park2020,Kawada2021}, etc.  
In this work, we study interplay between vorticity and spin currents where topological defects behave as positive and negative charges in the presence of electric and magnetic fields. We show that the steady state vorticity current induces the spin Hall current which can be measured by the inverse spin Hall or inverse magnetic spin Hall effects. We study this effect across the BKT transition analytically, and numerically using spin dynamics simulations. 

\begin{figure}
\centering
\includegraphics[width=0.8\linewidth]{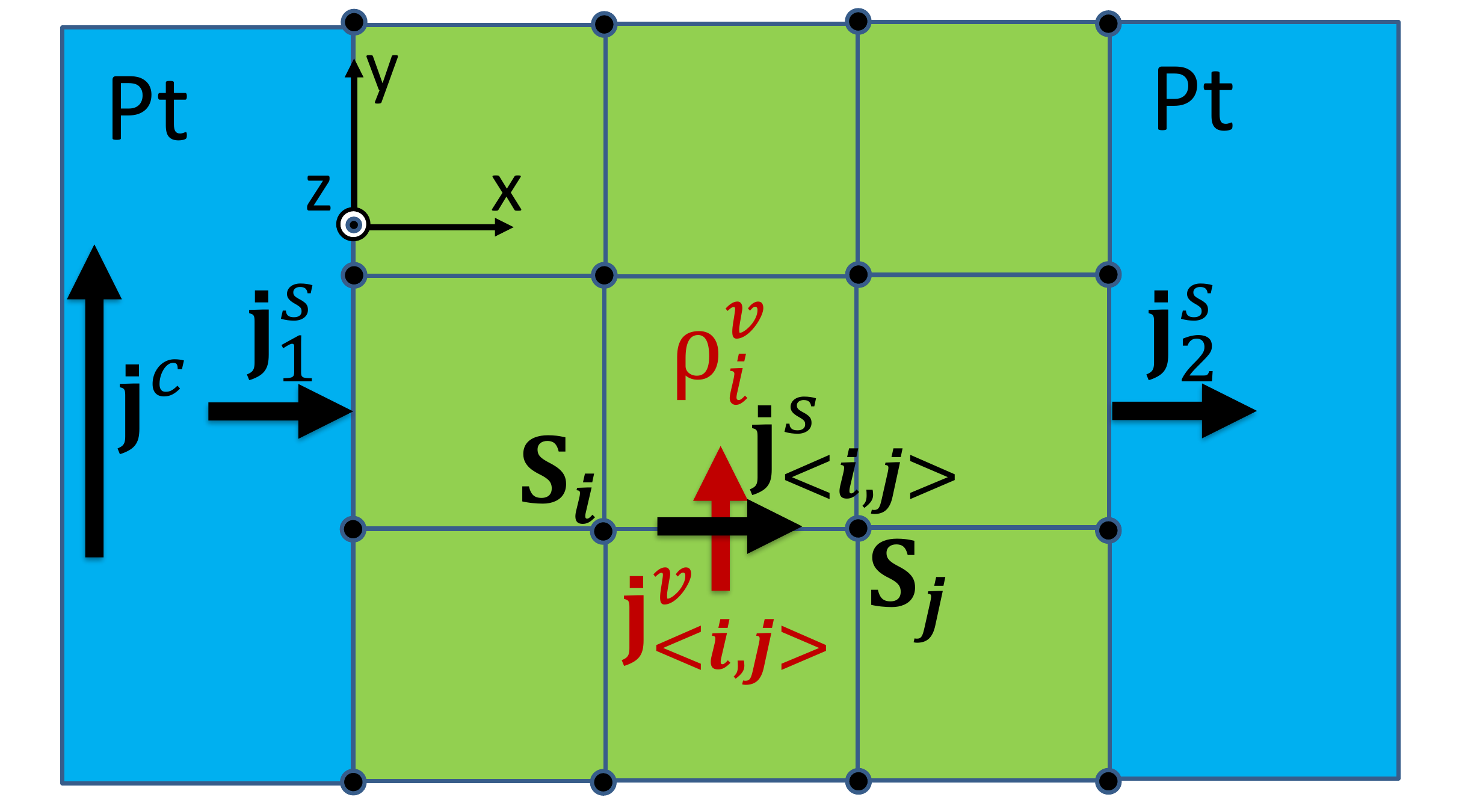}
\caption{(Color online) The vorticity current in an easy-plane magnet will lead to transverse spin Hall current that can be detected using a heavy metal such as Pt where $j^c$ is the direction of charge current. For a square lattice, spin and vorticity currents, and vorticity density are shown for a single plaquette.}
\label{fig:she}
\end{figure}

\textit{LLG dynamics}---We begin by considering the Hamiltonian describing a 2D magnetic insulator with the easy-plane magnetic anisotropy,
\begin{equation} \label{eq:ham}
    {\cal H} =-J \sum_{\left< i,j \right>} (S^x_i S^x_j+S^y_i S^y_j+\lambda S^z_i S^z_j)-2 J \beta \sum_i (S^z_i)^2 ,
\end{equation}
where $J<0$~\cite{Note} describes the antiferromagnetic exchange coupling, $\lambda$ ($0\leq\lambda<1$) describes the exchange anisotropy, and $\beta$ describes the single-ion magnetic anisotropy. We assume a square lattice; however, the approach also works for other lattices. The Hamiltonian~\eqref{eq:ham} can be realized in 2D vdW magnets~\cite{PhysRevLett.124.017201}. Depending on the strength of anisotropy, Eq.~\eqref{eq:ham} can lead to realizations of either in-plane vortices or merons~\cite{PhysRevB.39.11840}. We concentrate on the small anisotropy case realizing merons. The dynamical equations corresponding to the Hamiltonian~\eqref{eq:ham} can be readily obtained. With added dissipation and spin-orbit torque these lead to the discretized Landau-Lifshitz-Gilbert (LLG) equation,
\begin{equation}\label{eq:LLG}
     s(1+\alpha {\bf S}_i\times)\,\partial_t {\bf S}_i={\bf S}_i\times {\bf H}_i+\boldsymbol \tau_i^{so}. 
\end{equation}
Here $s$ stands for the spin density, $\alpha$ stands for the Gilbert damping, $\boldsymbol \tau_i^{so}$ is the spin-orbit torque due to spin Hall or magnetic spin Hall effects, the field ${\bf H}_i={\bf H}_i^\text{eff}+{\bf H}_i^\text{th}$ contains the effective field ${\bf H}_i^\text{eff}=\frac{J}{a^2}[\sum_{j\in N(i)} (S^x_j,S^y_j,\lambda S^z_j)+ (0,0,4\beta S^z_i)]$ with $N(i)$ denoting the nearest neighbours, and the thermal field ${\bf H}_i^\text{th}$ due to the Langevin force.
\begin{figure}
\centering
\includegraphics[width=0.7\linewidth]{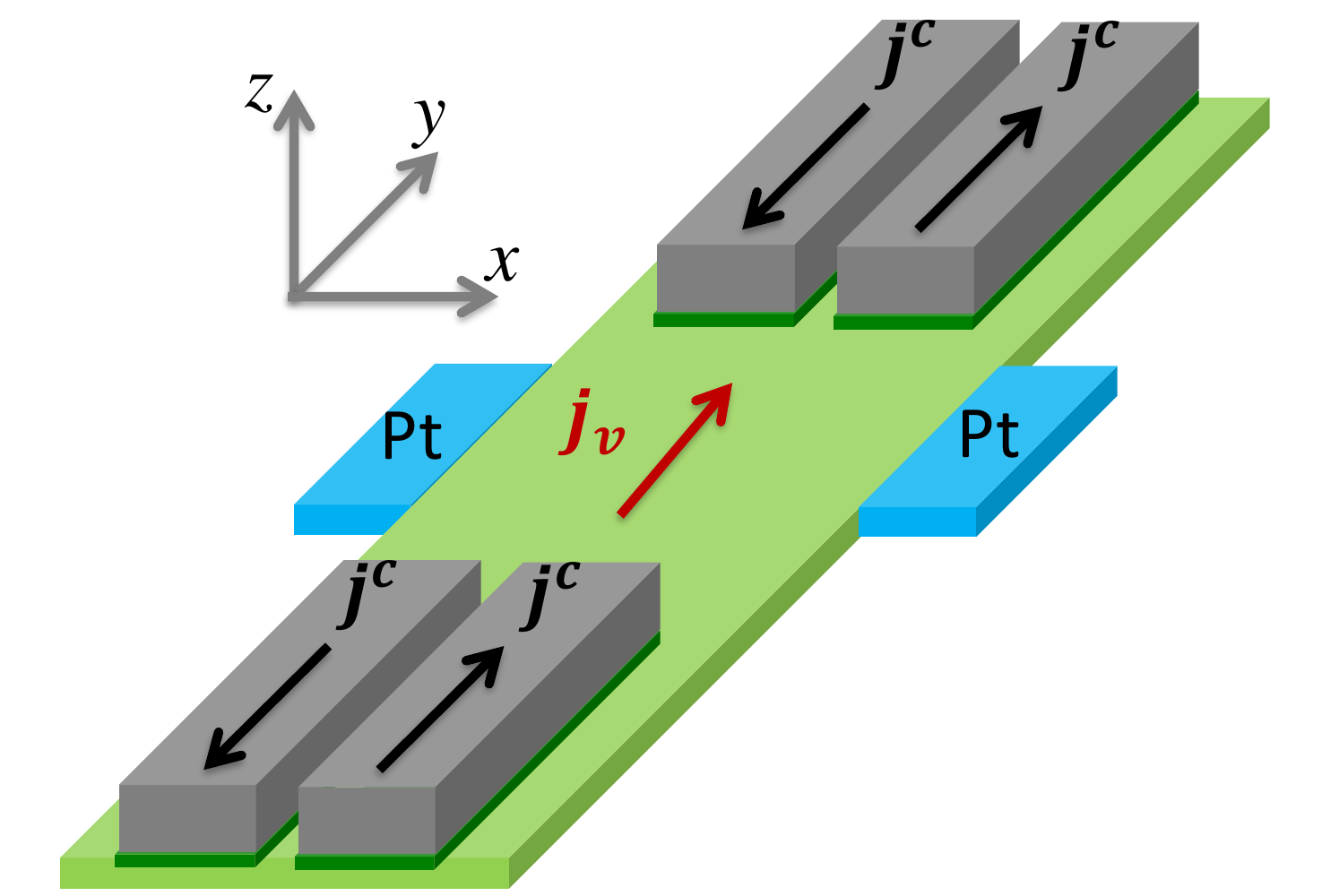}
\caption{(Color online) Two pairs of magnets with opposite charge currents $j^c$ are used to inject opposite spin currents with perpendicular polarization into an easy-plane magnet via the magnetic spin Hall effect~\cite{PhysRevResearch.2.023065,Kondou2021} or the unconventional spin-orbit torque~\cite{PhysRevB.102.014401,MacNeill2016}, establishing a steady vorticity current. The spin Hall effect of vorticity can be detected by injection of the spin current into heavy metal contacts, e.g., Pt contacts. The circuit can also run in reverse, where the injection of spin current through heavy metal contacts is used to generate the vorticity current and the inverse magnetic spin Hall effect is used to detect the spin current.}
\label{fig:circuit}
\end{figure}
The out-of-plane component of Eq.~\eqref{eq:LLG} can be rewritten in a form reminiscent of the continuity equation for spin current,
\begin{equation}\label{eq:spin}
    \partial_t \rho^s_i +{\boldsymbol \nabla}_i \cdot {\bf j}^s_{\left< i,j \right>}=-\alpha s\,{\bf z}\cdot{\bf S}_i\times\partial_t{\bf S}_i+{\bf z}\cdot\boldsymbol \tau_i^{so},
\end{equation}
where ${\boldsymbol \nabla}_i \cdot {\bf f}_{\left< i,j \right>}=\frac{1}{a^2}\sum_{j\in N(i)}({\bf r}_i-{\bf r}_j){\bf f}_{\left< i,j \right>}$ is the discrete divergence, $\rho^s_i=s S_i^z$ is the out-of-plane spin density, and ${\bf j}^s_{\left< i,j \right>}=\frac{J}{a^2}({\bf r}_i-{\bf r}_j) ({\bf z}\cdot{\bf S}_i\times {\bf S}_j)$ is the spin current~\cite{PhysRevB.100.144416} (see Fig.~\ref{fig:she}).  We can rewrite Eq.~\eqref{eq:spin} in the long wavelength limit as $\partial_t \rho^s+\boldsymbol \nabla {\bf j}^s=-\rho^s/\tau+{\bf z}\cdot\boldsymbol \tau^{so}$ for the dynamics in the vicinity of the in-plane configuration as follows from a relation, ${\bf z}\cdot{\bf S}\times\partial_t{\bf S}\propto \rho^s$ where $\tau$ has the meaning of the spin relaxation time. The spin density $\rho^s$ and the spin flux ${\bf j}^s$ form a spin three-current $\sigma^\mu=(\rho^s,{\bf j}^s)$. 

\textit{Electrodynamics of vorticity and dissipation}---The dynamics of the easy-plane magnet is influenced by topological defects. The conserved vorticity topological charge can be defined as
\begin{equation}\label{eq:charge}
    Q=\sum_{i\in S} \rho^v_i a^2=\frac{1}{2 \pi J}\ointctrclockwise_{\partial S} {\bf j}^s_{\left< i,j \right>} d{\bf l},
\end{equation}
where $\partial S$ is the boundary defined by a set of bonds forming a closed path and we introduce the vorticity density $\rho^v_i=\sum_{\left< i,j \right>\in {\cal P}(i)} \frac{1}{2 \pi a^2}({\bf z}\cdot{\bf S}_i\times {\bf S}_j) $ with ${\cal P}(i)$ describing all edges of plaquette $i$ with the counterclockwise ordering of edge indices $(i,j)$ (see Fig.~\ref{fig:she}). The right-hand side of Eq.~\eqref{eq:charge} can be seen as the consequence of the Stokes theorem. By identifying the spin current with the fictitious electric field~\cite{PhysRevLett.124.157203}, i.e., ${\bf E}= {\bf j}^s\times {\bf z}$, we can recast Eq.~\eqref{eq:charge} as the Gauss's law, $\int_{\partial S} d{\bf s}\cdot {\bf E}=2\pi J Q$. Note that the fictitious electric field, ${\bf E}_c=-\boldsymbol \nabla V$, corresponding to the electrostatic potential $V$ due to the two-dimensional Coulomb interaction of static topological defects~\cite{Kosterlitz_1974} will result in conserved equilibrium spin current. In the following discussion, we will use the spin current ${\bf j}^s$ and the electric field ${\bf E}$ interchangeably.

Dissipation associated with the Gilbert damping in Eq.~\eqref{eq:spin} in the presence of vorticity currents will lead to non-conserved spin currents. Injection of spin into 2D magnet, e.g., due to spin-orbit torque term in the right-hand side of Eq.~\eqref{eq:LLG}, 
\begin{equation}\label{eq:SO}
\boldsymbol \tau_i^{so}\propto {\bf S}_i\times[{\bf S}_i\times{\bf z}],
\end{equation}
will also lead to spin non-conservation. Such torques are typical for Pt contacts but in case of unconventional spin Hall effect will require materials with lower symmetry~\cite{PhysRevB.102.014401,MacNeill2016}.
The conserved vorticity current is defined for each bond as~\cite{PhysRevB.99.180402} ${\bf j}^v_{\left< i,j \right>}=\frac{1}{2 \pi a^2}{\bf z}\times({\bf r}_i-{\bf r}_j) [{\bf z}\cdot({\bf S}_i-{\bf S}_j)\times \partial_t ({\bf S}_i+{\bf S}_j)]$ (see Fig.~\ref{fig:she}). By direct inspection, we can confirm the following relation
\begin{equation}\label{eq:vcurrent}
    {\bf j}^v_{\left< i,j \right>}=\frac{1}{2 \pi J} {\bf z}\times\left(\partial_t {\bf j}^s_{\left< i,j \right>}+J\boldsymbol\nabla_{\left< i,j \right>}[{\bf z}\cdot{\bf S}\times\partial_t{\bf S}]\right),
\end{equation}
which establishes a relation between the vorticity current and spin dissipation due to the Gilbert damping in Eq.~\eqref{eq:spin}. Here $\boldsymbol \nabla_{\left< i,j \right>}f=(f_i-f_j)({\bf r}_i-{\bf r}_j)/a^2$ and it can be straightforwardly generalized to $\boldsymbol \nabla$ in the long wavelength limit.

The constitutive relations \eqref{eq:spin}, \eqref{eq:charge} and \eqref{eq:vcurrent} correspond to dissipative electrodynamics of vorticity which manifests after identification of spin three-current with the fictitious magnetic, $B=\rho^s$, and electric, ${\bf E}= {\bf j}^s\times {\bf z}$, fields~\cite{PhysRevLett.124.157203} acting on charged vortices where the charge corresponds to the vorticity topological number.

\textit{Vorticity and spin currents}---To drive the spin Hall response, we assume that a steady vorticity flow is present in the system. In Fig.~\ref{fig:circuit}, the vorticity flow is generated by injection of spin currents with opposite out-of-plane polarizations via the magnetic spin Hall effect~\cite{PhysRevResearch.2.023065,Kondou2021} or the unconventional spin-orbit torque~\cite{PhysRevB.102.014401,MacNeill2016}, e.g., by using additional layers covering the two-dimensional magnet through a nonmagnetic spacer. The vorticity current is maintained by the electric field ${\bf E}= {\bf j}^s\times {\bf z}$, i.e.,  
\begin{equation}\label{eq:response}
     {\bf j}_v=\sigma {\bf E},
\end{equation}
where $\sigma=\mu n_f$ is the conductivity expressed in terms of the mobility, $\mu$, and the total density of free topological
defects with positive and negative vorticity, $n_f$.

In a steady state, we combine Eqs.~\eqref{eq:spin}, \eqref{eq:vcurrent}, and \eqref{eq:response} and obtain
\begin{equation}\label{eq:transport}
    \frac{1}{2 \pi \alpha s} \boldsymbol\nabla (\boldsymbol\nabla\cdot {\bf j}^s-{\bf z}\cdot\boldsymbol \tau^{so})=\mu n_f {\bf j}^s, 
\end{equation}
where the total density of free topological defects $n_f$ behaves differently below and above the BKT transition. As a result, the system response described by Eq.~\eqref{eq:transport} also changes across the BKT transition. At $T>T_\text{BKT}$, Eq.~\eqref{eq:transport} is linear in ${\bf j}^s$ and the response is dominated by the temperature dependence of the free topological defect density, i.e., $ n_f \propto \exp(-2b/\sqrt{T/T_{BKT}-1})$~\cite{PhysRevLett.40.783}. At $T<T_\text{BKT}$, the free topological defects are absent in equilibrium. The spin current ${\bf j}^s$ breaks some bound pairs of topological defects as follows from Eq.~\eqref{eq:response} leading to the finite density~\cite{SciPostPhys} $n_f\propto \exp(-\Delta F/k_B T)$ with $\Delta F \approx \pi \Tilde{K} \ln({\cal J}^s/j^s)$ being the free energy barrier for unbinding a pair and ${\cal J}^s$ being a phenomenological parameter~\cite{Halperin2010,SciPostPhys}. This leads to non-linearity with respect to ${\bf j}^s$ in the right-hand side of Eq.~\eqref{eq:transport} as $n_f\propto (j^s/{\cal J}^s)^{\pi \Tilde{K}/k_B T}$ where $\Tilde{K}$ is the
vortex-renormalized spin stiffness.

\textit{Spin Hall effect}---We consider the vorticity flow in Fig.~\ref{fig:circuit}, and show that this results in spin current injection into Pt contacts. To this end, we consider a region of an easy-plane magnet in Fig.~\ref{fig:she} and assume a steady vorticity current. As pushing the vorticity through a magnetic film results in transverse winding~\cite{PhysRevB.102.224433}, this should also lead to spin current pumping into Pt contacts in Fig.~\ref{fig:she}. Assuming that the Pt contacts in Fig.~\ref{fig:she} are only used as detectors, we arrive at the following boundary conditions
\begin{align}\label{eq:boundary1}
    j^s(0)&=j^s_1+\Tilde{g}\partial_x j^s |_{x=0},\\ 
     j^s(L_x)&=j^s_2-\Tilde{g}\partial_x j^s |_{x=L_x}, \label{eq:boundary2}
\end{align}
where $\Tilde{g}=a \alpha^\prime/\alpha$ describes the increased Gilbert damping $\alpha^\prime$ (in general different for the left and right boundaries in Fig.~\ref{fig:she}) at $x=0$ and $x=L_x$ due to contact with the detector~\cite{PhysRevResearch.4.023236}. The parameters $j^s_1$ and $j^s_2$ describe the spin current at the boundary due to vorticity flow; for symmetrical setup $j^s_1=j^s_2=j^s$. Assuming that $\Tilde{g}/L_x$ is small, we can further approximate the spin currents injected into heavy metal contacts in Fig.~\ref{fig:she} as $j^s(0)-j^s_{\Tilde{g}=0}=\Tilde{g}\partial_x j^s_{\Tilde{g}=0} |_{x=0}$ and $j^s(L_x)-j^s_{\Tilde{g}=0}=-\Tilde{g}\partial_x j^s_{\Tilde{g}=0} |_{x=L_x}$ where $j^s_{\Tilde{g}=0}$ is the spin current calculated for boundaries with $\Tilde{g}=0$~\cite{PhysRevResearch.4.023236}.

In the symmetrical setup, the analytical solutions of Eq.~\eqref{eq:transport} can be obtained using the inverse function $j(x)=f^{-1}(|x-L_x/2|)$ in terms of the reduced spin current $j = j^s/{\cal J}^s $ as
\begin{align}\label{eq:spin-current}
    f(j)=-\frac{j \sqrt{\nu \lambda^2}}{(j_0)^{\nu}} \operatorname{Im}\left[ \prescript{}{2}{F}_1\Big( \frac{1}{2}, \frac{1}{2\nu}, 1 + \frac{1}{2\nu}; \frac{j^{2\nu}}{j_0^{2\nu}}\Big)\right] ,
\end{align}
where $\prescript{}{2}{F}_1(\dots)$ stands for the hypergeometric function, $\nu=1+\pi\Tilde{K} /2 k_B T$, and $\lambda^2=2\pi \alpha s \mu n_f(j^s/{\cal J}^s)^{\pi\Tilde{K} /k_B T}$. The boundary condition $j(0)=j(L_x)=j^s_{\Tilde{g}=0}/{\cal J}^s$ can be used to find the unknown constant $j_0$ equal to the reduced spin current density at $x=L_x/2$. For $\Tilde{K}>0$, Eq.~\eqref{eq:spin-current} describes algebraic decay of spin current from edges up to $x=L_x/2$. For $\Tilde{K}=0$, ${\cal J}^s$ cancels out and Eq.~\eqref{eq:transport} corresponds to the spin diffusion where the solution in Eq.~\eqref{eq:spin-current} describes exponential decay away from the edges. 
We introduce the averaged vorticity current according to relation
\begin{equation}\label{eq:Vcurrent}
    J^v=\frac{1}{L_x}\int_0^{L_x} j^v dx .
\end{equation}
 It is convenient to use $j^s_{\Tilde{g}=0}$ to characterize the vorticity bias as follows from its relations to the averaged vorticity current.
 The averaged vorticity current for $T<T_\text{BKT}$ becomes
\begin{align}\label{eq:nonlinear1}
    J^v=\frac{{\cal J}^s}{\pi\alpha s }\frac{\sqrt{(j^s_{\Tilde{g}=0}/{\cal J}^s)^{2\nu}-(j_0)^{2\nu}}}{\sqrt{\nu} \lambda L_x}
    &\stackrel{\frac{j L_x }{\lambda}\gg 1}{\approx}\frac{{\cal J}^s}{\pi\alpha s} \frac{(j^s_{\Tilde{g}=0}/{\cal J}^s)^{\nu}}{\sqrt{\nu}\lambda L_x}\,,\\
    &\stackrel{\frac{j L_x}{\lambda}\ll 1}{\approx}\frac{{\cal J}^s}{\pi\alpha s} \frac{(j^s_{\Tilde{g}=0}/{\cal J}^s)^{2\nu-1}}{2(\lambda)^2}\,,\label{eq:nonlinear2}
\end{align}
and for $T>T_\text{BKT}$,
\begin{align}\label{eq:vort-tot}
    J^v=\frac{j^s_{\Tilde{g}=0}}{\pi\alpha s }\frac{1}{\lambda L_x\coth(L_x/2\lambda)}
    &\stackrel{\frac{L_x }{\lambda}\gg 1}{\approx}\frac{1}{\pi\alpha s} \frac{j^s_{\Tilde{g}=0}}{\lambda L_x}\,,\\
    &\stackrel{\frac{L_x }{\lambda}\ll 1}{\approx}\frac{1}{\pi\alpha s} \frac{j^s_{\Tilde{g}=0}}{2(\lambda)^2}\,.
\end{align}
Above relations show that the averaged vorticity current scales linearly with $j^s_{\Tilde{g}=0}$ above the BKT transition. Below the BKT transition, the scaling will be nonlinear with the power factors given in Eqs.~\eqref{eq:nonlinear1} and ~\eqref{eq:nonlinear2}.
\begin{figure}
\centering
\includegraphics[width=0.9\linewidth]{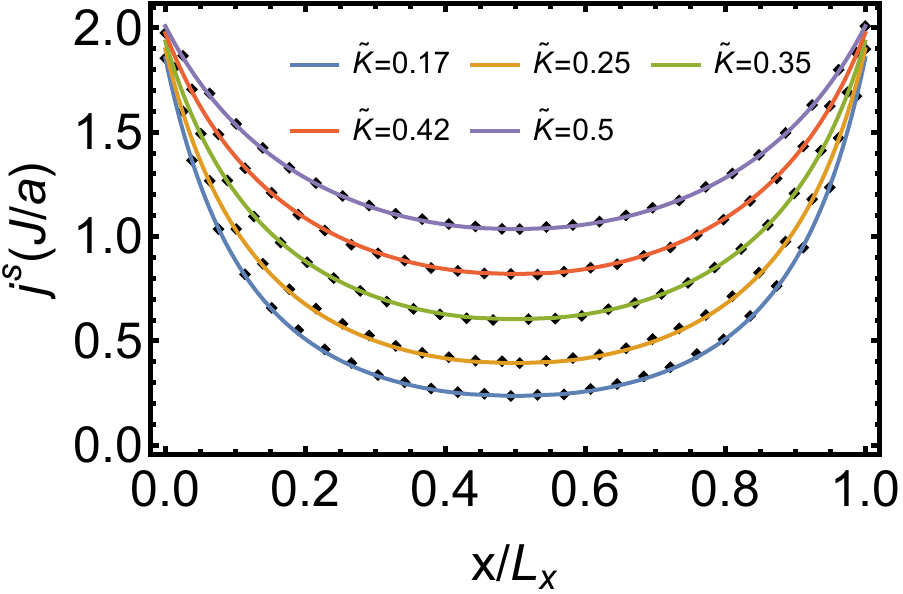}
\caption{(Color online) The spin current density in response to a steady vorticity flow. The spin current can be detected by the inverse spin Hall effect or the inverse magnetic spin Hall effect in a contact. The temperature varies across the BKT transition which is reflected by the values of the spin stiffness $\Tilde{K}$. The lines correspond to analytical results, and diamond symbols represent numerical calculations using the spin dynamics simulations with the in-plane magnetic anisotropy described by $\beta=0.05$.}
\label{fig:spin-current}
\end{figure}

Using results for the vorticity current, we can write the spin currents
injected into heavy metal contacts as
\begin{equation}\label{eq:resp}
    j^s(0)-j^s_{\Tilde{g}=0}=j^s(L_x)-j^s_{\Tilde{g}=0}=(a^2 s)\theta_{sh}  J^v,
\end{equation}
where $\theta_{sh}=-\pi\alpha^\prime (L_x/a)$ has the meaning of the spin Hall angle. This response is detectable by heavy metal contacts via inverse spin Hall effect as shown in Fig.~\ref{fig:she}.
\begin{figure}
\centering
\includegraphics[width=\linewidth]{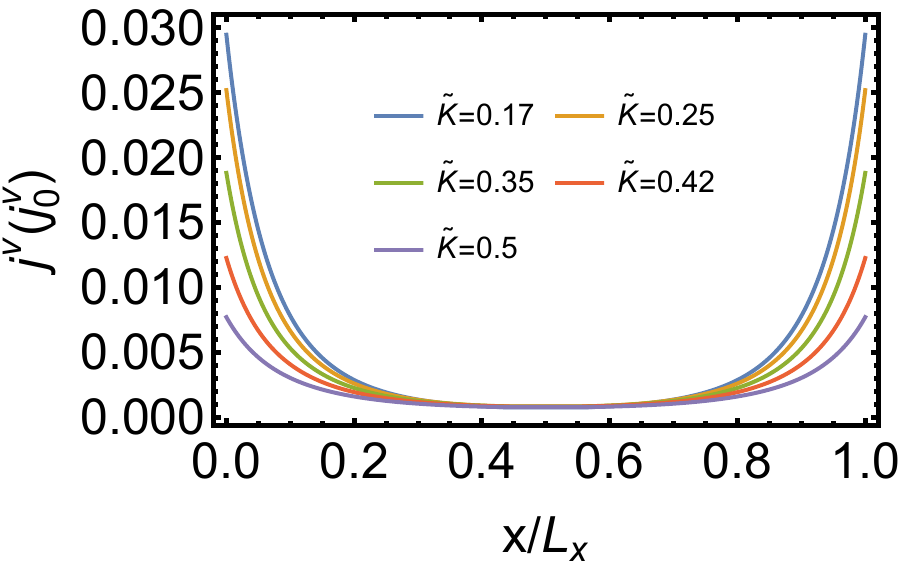}
\caption{(Color online) The distribution of the vorticity current density expressed in units of $j_0^v=J/(2\pi\alpha s a^3)$. The temperature varies across the BKT transition which is reflected by the values of the spin stiffness $\Tilde{K}$. We use the in-plane magnetic anisotropy described by $\beta=0.05$.}
\label{fig:v-current}
\end{figure}

\textit{Numerical results}---
We performed spin dynamics simulations of lattice described by Hamiltonian~\eqref{eq:ham} with $1600\times200$ sites and periodic boundary conditions.
Calculated quantities are averaged over time as well as $100$ initial configurations. We consider the in-plane magnetic anisotropy induced either by anisotropic exchange or by single-ion magnetic anisotropy and obtain similar results for both cases. The Gilbert damping and the Langevin random force are calculated for $\alpha=0.0005$. We obtain averaged quantities $\left< j^s \right>_\text{av}$ and $ \left< \rho^s \right>_\text{av}$ after the steady state is established. We use the feedback-optimized
parallel tempering Monte Carlo~\cite{Katzgraber2006,PhysRevResearch.4.023236} with $10^6$ metropolis updates per spin to obtain $100$ initial configurations for spin dynamics simulations. 
\begin{figure}
\centering
\includegraphics[width=0.9\linewidth]{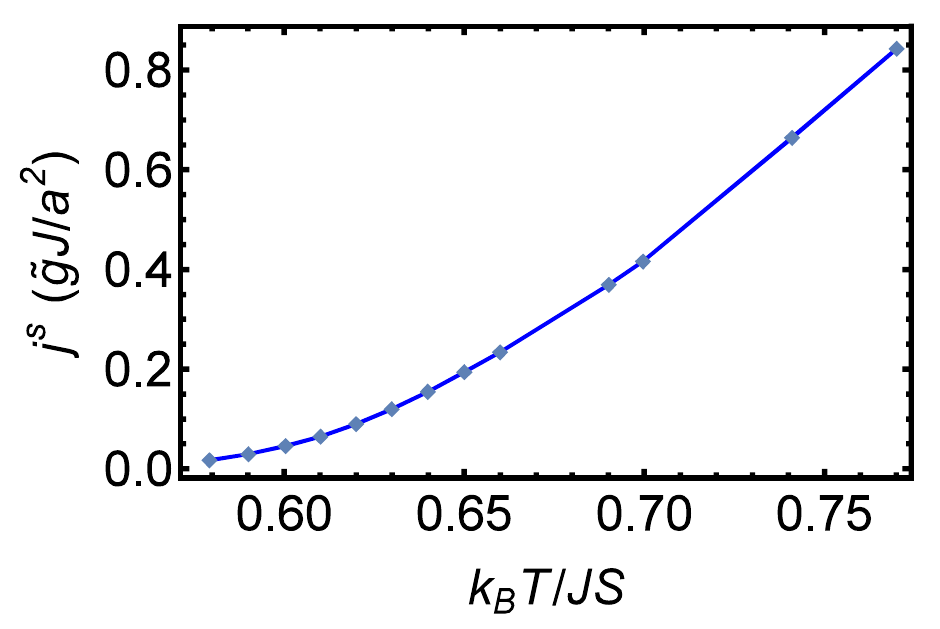}
\caption{(Color online) The diamond symbols represent the measurable spin Hall response in Eq.~\eqref{eq:resp} obtained by the spin dynamics simulations. }
\label{fig:spin-out}
\end{figure}

The spin is injected at $x=0$ and absorbed at $x=L_x$ to mimic the effect of $j^s_{\Tilde{g}=0}$ in boundary conditions in Eqs.~\eqref{eq:boundary1} and \eqref{eq:boundary2}. We take $\Tilde{g}=0$ as this term does not lead to substantial changes when it is small. In Fig.~\ref{fig:spin-current}, we plot the profile of spin current density along the x axis as the temperature is varied across the BKT transition, which is indicated by the values of the spin stiffness. Note that for an infinite system one expects the spin stiffness jump from $\Tilde{K}=2 k_B T_\text{BKT}/\pi$ to zero at the BKT transition while finite size effects can lead to more gradual changes in the spin stiffness~\cite{PhysRevB.49.12071,PhysRevResearch.4.023236}. We observe good agreement between the fitted analytical results of Eq.~\eqref{eq:spin-current} and the numerical results shown by diamond symbols. For smaller spin stiffness, we observe faster (but still algebraic corresponding to spin superfluidity) decay of spin current in the vicinity of boundaries. At high enough temperature, $\Tilde{K}=0$ and the decay becomes exponential (not shown). In Fig.~\ref{fig:v-current}, we plot the profile of the vorticity current density obtained from the fitted spin current in Fig.~\ref{fig:spin-current}. The averaged vorticity current given by Eq.~\eqref{eq:Vcurrent} increases as the spin stiffness decreases to zero around $k_B T/J S\approx 0.68 $. At the same time the vorticity current density decays faster in the vicinity of edges as the spin stiffness decreases. The exponential decay is recovered when $\Tilde{K}=0$. In Fig.~\ref{fig:spin-out}, we plot the spin current response in Eq.~\eqref{eq:resp} obtained by spin dynamics simulations. The increase in the spin current is consistent with the behavior of the free topological defect density above the BKT transition, i.e., $ n_f \propto \exp(-2b/\sqrt{T/T_{BKT}-1})$~\cite{PhysRevLett.40.783}.

\textit{Conclusions}---We studied the interplay of spin and vorticity currents where topological defects play a role of positive and negatie charges in the presence of fictitious electric and magnetic fields. We showed that the steady state vorticity current can induce the spin Hall current which can be measured by the inverse spin Hall or inverse magnetic spin Hall effects. The proposed effect is opposite to spin injection induced vorticity flow discussed in Ref.~\cite{PhysRevB.102.224433}. We studied the effect across the BKT transition analytically, and numerically using spin
dynamics simulations. We point-out that to define vorticity circuit one can use an easy-plane magnet below the BKT transition while the conducting channel can be realized via p- or n-doping (excess of positive or negative vorticity), e.g., using the field-like torque suggested in Ref.~\cite{PhysRevB.99.180402}, 
$\boldsymbol \tau=\gamma S_z ({\bf j}\cdot\boldsymbol\nabla){\bf S}$, induced by current ${\bf j}$ 
in perpendicularly magnetized magnet which is in contact with an easy-plane magnet. Note that the same torque can also be represented as the Dzyaloshinskii-Moriya interaction (DMI) with the DMI vector pointing out of plane. Our results demonstrate possibilities for low power computing and logic devices relying on vorticity and spin flows.

We acknowledge useful discussions with K. Belashchenko. This work was supported by the U.S. Department of Energy, Office of Science, Basic Energy Sciences, under
Award No. DE-SC0021019. Part of this work was completed utilizing the Holland Computing Center of the
University of Nebraska, which receives support from the Nebraska Research Initiative.

\bibliography{lib}

\begin{thebibliography}{57}
\bibitem [{Not()}]{Note}%
  \BibitemOpen
  \href@noop {} {}\bibinfo {note} {For $J<0$, we obtain qualitatively similar
  results.}\BibitemShut {Stop}%
\end{thebibliography}

\end{document}